\documentclass{article}
\usepackage{ijcai15}
\usepackage{amsfonts,amssymb,mathtools,amsthm}
\usepackage{subcaption}
\usepackage{graphicx, xcolor}
\usepackage{times}
\usepackage{amsfonts}
\usepackage{algorithm,algorithmic}
\usepackage{multirow}

\title{Social Trust Prediction via Max-norm Constrained 1-bit Matrix Completion}

\author{
Jing Wang$^1$, Jie Shen$^{2}$, Huan Xu$^{3}$\\
$^1$ 
 Hefei University of Technology, China\\
wangjing@mail.hfut.edu.cn\\
$^2$ 
Rutgers University, USA\\ js2007@rutgers.edu \\
$^3$ 
National University of Singapore, Singapore\\ mpexuh@nus.edu.sg \\
}
\numberwithin{equation}{section}

\theoremstyle{plain}
\newtheorem{theorem}{Theorem}[section]
\newtheorem{lemma}[theorem]{Lemma}

\theoremstyle{definition}

\theoremstyle{remark}

\renewcommand{\(}{\left(}
\renewcommand{\)}{\right)}

\newcommand{\lv}{\left\vert}
\newcommand{\rv}{\right\vert}

\newcommand{\st}{\textrm{s.t.}}

\newcommand{\bu}{\mathbf{u}}
\newcommand{\bv}{\mathbf{v}}

\newcommand{\bm}{\mathbf{m}}

\newcommand{\Rpd}{\mathbb{R}^{p\times d}}
\newcommand{\Rpp}{\mathbb{R}^{p\times p}}

\newcommand{\Rnd}{\mathbb{R}^{n\times d}}
\newcommand{\Rpn}{\mathbb{R}^{p\times n}}
\newcommand{\Rnn}{\mathbb{R}^{n\times n}}

\newcommand{\mP}{\mathcal{P}_{\Omega}}

\newcommand{\twonorm}[1]{\left\lVert #1 \right\rVert_{2}}

\newcommand{\twoinfnorm}[1]{\left\lVert #1 \right\rVert_{2,\infty}}
\newcommand{\maxnorm}[1]{\left\lVert #1 \right\rVert_{\max}}
\newcommand{\nuclearnorm}[1]{\left\lVert #1 \right\rVert_{*}}
\newcommand{\fronorm}[1]{\left\lVert #1 \right\rVert_{F}}

\newcommand{\trans}{^\top}

\begin{document}

\maketitle

\begin{abstract}
Social trust prediction addresses the significant problem of exploring interactions among users in social networks. Naturally, this problem can be formulated in the matrix completion framework, with each entry indicating the trustness or distrustness. However, there are two challenges for the social trust problem: 1) the observed data are with sign (1-bit) measurements; 2) they are typically sampled non-uniformly. Most of the previous matrix completion methods do not well handle the two issues. Motivated by the recent progress of max-norm, we propose to solve the problem with a 1-bit max-norm constrained formulation. Since max-norm is not easy to optimize, we utilize a reformulation of max-norm which facilitates an efficient projected gradient decent algorithm. We demonstrate the superiority of our formulation on two benchmark datasets.
\end{abstract}

\section{Introduction}
\label{sec:intro}
With the increasing popularity of social networks, there exist many interesting and difficult problems, such as friends recommendation, information propagation, etc. 
 In this paper, we study the problem of social trust prediction, which aims to estimate the positive or negative relationship among the users based on 
 the existing trustness information associated with them. 
 This problem plays an important role in social networks as the system can block the invitation from someone that the user does not trust,  or recommend new friends who enjoys a high reputation.

Naturally, the social trust problem can be formulated within the matrix completion framework~\cite{liben2007link} \cite{billsus1998learning}. That is, the $(i, j)$-th entry of the observed data matrix $Z \in \Rnn$ is a 1-bit code implying that the $i$-th user trusts the $j$-th user if $Z_{ij}= 1$. Here, $n$ denotes the number of users. However, what we observe is only a small fraction of the entries, whose values are zero. And our goal is to estimate the missing entries according to the 1-bit measurements in $Z$.

Note that the problem is ill-posed if no assumption is imposed on the structure of the data. To solve the problem, a number of methods are proposed. Generally, existing social trust prediction methods fall into three categories. The first category is based on similarity measures or the structural context similarity ~\cite{newman2001clustering} \cite{chowdhury2010introduction} \cite{katz1953new} \cite{jeh2002simrank}, motivated by the intuition that an individual tends to trust their neighbors, or the ones with similar trusted people. The second is based on low rank matrix completion~\cite{billsus1998learning} \cite{cai2010singular} \cite{huang2013social}, which assumes that the underlying data matrix is low-rank or can be approximate by a low-rank matrix. The third one models the problem as a binary classification one and utilizes techniques such as logistic regression \cite{leskovec2010predicting}. 

\textbf{Challenges.} However, there are two issues emerging in social trust prediction which are not well characterized by the algorithms in previous works. First, the value of the observed entry is either 1 or $-1$, which is analogous to the binary classification problem. But in our problem, we are handling much more complex matrix data. Fortunately,~\cite{srebro2004mmmf} presented a maximum margin matrix factorization framework that unifies the binary problem for vector case and matrix case. The key idea in their work is a low-norm matrix factorization, which will also be utilized in this paper. Second, the locations of the entries are sampled non-uniformly, which gaps the theory and practice for a lot of matrix completion algorithms. To tackle this challenge, we suggest using the max-norm as a convex surrogate for the rank function, which is shown to be superior to the well-known nuclear norm when addressing the non-uniform data~\cite{srebro2010non-uniform}.


\textbf{Our contributions} are two-folds: 1) To the best of our knowledge, we are the first to address the social trust prediction problem by utilizing a max-norm constrained formulation. 2) Although a max-norm constrained problem can be solved by SDP solvers and an accurate enough solution can be achieved, we here utilize a projected gradient algorithm that is scalable to large scale datasets. We empirically show the improvement of our formulation for the non-uniform 1-bit benchmarks compared to state-of-the-art solvers.

\section{Related Work}
\label{relat}
\begin{figure}[!t]
	\centering
	\includegraphics[width=0.5\textwidth]{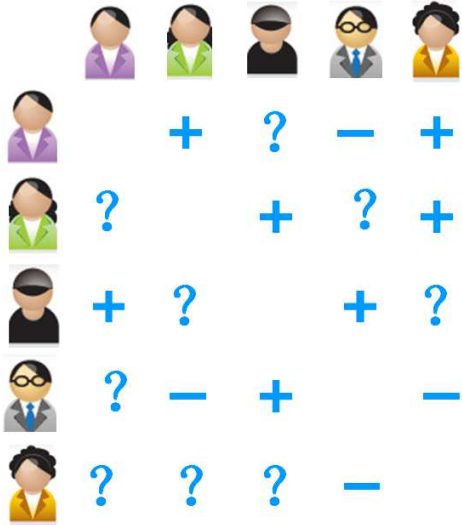}
		\vspace{-3em}
	\caption{Illustration of the data matrix of social trust. Each column of the data matrix is a rating sample. Each entry is a tag that a user assigns to another user. The symbol ``+'' denotes the ``trust'' relationship, ``-'' denotes ``distrust'' and ``?'' means unknown relationship (that we aim to predict). Most of the relationships are unknown, so the data is sparse. Typically, each individual has own preference and friendship network, making the data non-uniform. Also noted that the data matrix may not be symmetric.}
	\label{trusg}
		\vspace{-1em}
\end{figure}

Social interaction is investigated intensively in the last decades. The social interaction indicates the friendship, support, enemy or disapproval as shown in Figure 
\ref{trusg}. Online users rely on the trustworthiness information to filter information, establish collaboration or build social bonds. Social networks rely on the trust information to make recommendation, attract users from other circles, or influencing public opinions. Thus, the exploration of social trust has a wide range of applications, and has emerged as an important topic in social network research. A number of methods are proposed.

One kind of methods are based on the similarity measurement. Specifically, Jaccard's coefficient is commonly used to measure the probability two items that have a relationship. Inspired by the metric, Jeh and Widom~\cite{jeh2002simrank} proposed a domain-independent similarity measurement, SimRank. \cite{newman2001clustering} directly defined a score to verify the correlation between common neighbors. Some methods are based on relational data modeling, structural proximity measures and stochastic relational model \cite{getoor2005link} \cite{liben2007link} \cite{yu2009large}. The above mentioned methods are mainly derived from the solutions of link prediction. The link prediction is oriented to network-level prediction, whereas social trust prediction focuses on person-level. Another class of methods are derived from the collaborative filtering methods, such as clustering techniques \cite{sarwar2001item}, model-based methods \cite{hofmann1999latent}, and the matrix factorization models \cite{srebro2003weighted} \cite{mnih2007probabilistic}. However, the data matrix of trust has some structure properties different from the user-item matrix, such as transitivity. 
Meanwhile, the social trust in reality is extremely sparse. For instance, Facebook has hundreds of millions of users, but most of them have less than 1,000 friends. Besides, the people with similar personality tend to behave similarly. To sum up, the data matrix of social trust has both sparse and low-rank structure. Thus, the social trust prediction problem is especially suitable for the matrix completion  model. That is the focus of our paper.

The problem of matrix completion is to recover a low-rank matrix from a subset of entries \cite{candes2009exact}, which is given by:
\begin{equation}
	\label{eqRank}
	\begin{split}
		\min_X~ & \text{rank}(X) \\
		s.t. ~~&P_{\Omega}(X)= P_{\Omega}(Z)
	\end{split}
\end{equation}
where $Z\in \Rpn$ is the data matrix, $X$ is the recovered matrix, and $\Omega$ is the index set of observed entries. Th optimization problem~\eqref{eqRank} is not only NP-hard, but requires double exponential time complexity with the number of samples~\cite{recht2010guaranteed}. To solve the above problem, one alternative is to use nuclear norm as a relaxation to the rank function:
\begin{equation}
	\label{traceN}
	\begin{split}
		\min_X~ & ||X||_{\ast} \\
		s.t.~~&P_{\Omega}(X)= P_{\Omega}(M)
	\end{split}
\end{equation}
where $||X||_{\ast}$ denotes the sum of singular values of matrix $X$. \cite{cai2010singular} developed a first-order procedure to solve the convex problem (\ref{traceN}), namely singular value thresholding (SVT). \cite{jain2010guaranteed} minimized the rank minimization by the singular value projection (SVP) algorithm. \cite{keshavan2010matrix} solved the problem by first trimming each row and column with too few entries, then compute the truncated SVD of the trimmed matrix. Under certain conditions, it showed accurate recovery on the order of $nd \log n$ samples ($n$ is the number of samples, $d$ is the rank of recovered matrix). With the rapid development of matrix completion problem, some more efficient methods have been proposed \cite{candes2010power}\cite{gross2011recovering}\cite{wang2012stability}\cite{huang2013social}.

However, all the methods mentioned above use the nuclear norm as the surrogate to the rank, whose exact recovery can be guaranteed only when the data are sampled uniformly, which is not practical in real world applications. On the other hand, recent empirical on max-norm~\cite{srebro2004mmmf} shows promising results for non-uniform data if one utilize the max-norm as a surrogate~\cite{srebro2010non-uniform}. Notably, for some specific problems, such as collaborative filtering,~\cite{srebro2005rank} proved that the generalization error bound for max-norm is better than the nuclear norm. More recently,~\cite{shen2014online} reported encouraging results on the subspace recovery task (which is closely relevant to matrix completion). Since the social trust data is non-uniformly sampled, we believe that a max-norm regularized formulation can better handle the challenge than the nuclear norm. Our formulation is also inspired by a recent theoretical study on matrix completion with 1-bit measurement~\cite{cai2013max}, which established a minimax lower bound on the general sampling model and derived the optimal convergence rate in terms of Frobenius norm loss. Furthermore, there are several practical algorithms to solve max-norm regularized or max-norm constrained problems, see~\cite{lee2010practical} and~\cite{shen2014online} for example.

\subsection{Overview}
After review of related work in Section~\ref{relat}, we introduce the notations and formulate the problem in Section~\ref{sec:notation}. Then we give algorithm to solve the max-norm constrained 1-bit matrix completion (MMC) problem in Section~\ref{sec:alg}. Meanwhile, we also provide an equivalent SDP formulation for the MMC, which can be accurately solved at the expense of efficiency. Then we report the empirical study on two benchmark datasets in Section~\ref{sec:exp}. Section~\ref{sec:conclusion} concludes this paper and discusses possible future work.

\section{Notations and Problem Setup}
\label{sec:notation}
In this section, we introduce the notations that will be used in this paper. Capital letters such as $M$ are used for matrices and lowercase bold letters such as $\bv$ denotes vectors. The $i$-th row and $j$-th column of a matrix $M$ is denoted by $\bm(i)$ and $\bm_j$ respectively, and the $(i, j)$-th entry is denoted by $M_{ij}$. For a vector $\bv$, we use $v_i$ to denote its $i$-th element. We denote the $\ell_2$ norm of a vector $\bv$ by $\twonorm{\bv}$. For a matrix $M \in \Rpn$, we denote the Frobenius norm by $\fronorm{M}$ and $\twoinfnorm{M}$ denotes the maximum $\ell_2$ row norm of $M$, {\em i.e.,}
\begin{equation*}
\twoinfnorm{M} := \max_{i=1}^{p} \twonorm{\bm(i)}.
\end{equation*}
We further define the {\em max-norm} of $M$ \cite{linial2007complexity},
\begin{equation}
\label{eq:max def}
\maxnorm{M} = \min_{U, V, M=UV\trans} \max \{ \twoinfnorm{U}^2, \twoinfnorm{V}^2 \},
\end{equation}
where we enumerate all possible factorizations to obtain the minimum.

\textbf{Intuition on max-norm.} At a first sight, the max-norm is hard to understand. We simply explain why it is a tighter approximation to the rank function than the nuclear norm. Again, we write the nuclear norm of $M$ as a factorization form~\cite{recht2010guaranteed}:
\begin{equation*}
\nuclearnorm{M} := \min_{U, V, M=UV\trans} \frac{1}{2} \( \fronorm{U}^2 + \fronorm{V}^2 \).
\end{equation*}
Note that the Frobenius norm is the sum of the square of the $\ell_2$ row norm. Thus, a nuclear norm regularizer actually constrains the average of the $\ell_2$ row norm, while the max-norm constrains the maximum of the $\ell_2$ row norm!


Given the observed data $Z\in \Rpn$, we are interested in approximating $Z$ with a low-rank matrix $X$, which can be formulated by,
\begin{equation*}
\begin{split}
\text{minimize}\ & \frac{1}{2} \fronorm{\mP(Z-X)},\\
\st\ & \text{rank}(X) \leq d,
\end{split}
\end{equation*}
where $\Omega$ is an index set of observed entries and $d$ is some expected rank. $\mP(M)$ is a projection operator on a matrix $M$ such that $\mP(m_{ij}) = m_{ij}$ if $(i ,j)\in \Omega$ and zero otherwise. However, it is usually intractable to optimize the above program as the rank function is non-convex and non-continuous~\cite{recht2010guaranteed}. One common approach is to use the nuclear norm as a convex surrogate to the rank function. However, it is well known that the nuclear norm cannot well handle the non-uniform data. Motivated by the recent progress in max-norm~\cite{srebro2010non-uniform,cai2013max,shen2014online}, we use the max-norm  as an alternative convex relaxation, which gives the following formulation:
\begin{equation}
\label{eq:main prob}
\begin{split}
\min_X\ & \frac{1}{2}\fronorm{\mP(Z-X)}^2,\\
\st\ & \maxnorm{X} \leq \lambda^2,
\end{split}
\end{equation}
where $\lambda$ is some tunable parameter.

\section{Algorithm}
\label{sec:alg}
The max-norm is convex and moreover, it can be solved by any SDP solver. Formally, we have the following lemma:
\begin{lemma}[\cite{srebro2004mmmf}]
\label{lem:max}
For any matrix $X \in \Rpn$ and $\lambda \in \mathbb{R}$, $\maxnorm{X} \leq \lambda$ if and only if there exist $A \in \Rpp$ and $B \in \Rnn$, such that $\begin{bsmallmatrix}
A & X\\
X\trans & B
\end{bsmallmatrix}$
is semi-definite positive and each diagonal element in $A$ and $B$ is upper bounded by $\lambda$.
\end{lemma}

With Lemma~\ref{lem:max} on hand, one can formulate Problem~\ref{eq:main prob} as an SDP:
\begin{equation}
\label{eq:sdp prob}
\begin{split}
\min_{X,A,B}\ & \frac{1}{2}\fronorm{\mP(Z-X)}^2,\\
\st\ & A_{ii} \leq \lambda^2,\ B_{jj} \leq \lambda^2,\ \forall\ i \in [p],\ j \in [n],\\
& \begin{bmatrix}
A & X\\
X\trans & B
\end{bmatrix} \succeq 0.
\end{split}
\end{equation}
And this program can be solved by any SDP solver to obtain accurate enough solution.

However, SDP solvers are not scalable to large matrices. Thus, in this paper, we apply a projected gradient method to solve Problem~\eqref{eq:main prob}, which is due to~\cite{lee2010practical}. A key technique is the reformulation of the max-norm~\eqref{eq:max def}. Assume that the rank of the optimal solution $X^*$ produced by the SDP~\eqref{eq:sdp prob} is at most $d$. Then we can safely factorize $X = UV\trans$, with $U \in \Rpd$ and $V \in \Rnd$. Combining the factorization and the definition, we obtain the following equivalent program 
:
\begin{equation}
\label{eq:uv prob}
\begin{split}
\min_{U,V}\ & \frac{1}{2}\fronorm{\mP(Z-UV\trans)}^2,\\
\st\ & \twoinfnorm{U} \leq \lambda,\ \twoinfnorm{V} \leq \lambda.
\end{split}
\end{equation}
Note that the gradient of the objective function w.r.t. $U$ and $V$ can be easily computed. That is,
\begin{equation}
\label{eq:gradient}
\begin{split}
\nabla_U^{} f(Z, U, V) &= \mP\((UV\trans - Z)V \),\\
\nabla_V^{} f(Z, U, V) &= \mP\( (VU\trans - Z\trans)U \).
\end{split}
\end{equation}
Here, for simplicity we define
\begin{equation*}
f(Z, U, V) = \frac{1}{2}\fronorm{\mP(Z-UV\trans)}^2.
\end{equation*}
The inequality constraints can be addressed by a projection step. That is, when we have a new iterate $(U_t, V_t)$ at the $t$-th iteration, we can check if they violate the constraints. If not, we can proceed to the next iteration. Otherwise, we can scale the rows of $U$ and/or $V$ by $\frac{\lambda}{\twoinfnorm{U}}$ and/or $\frac{\lambda}{\twoinfnorm{V}}$ respectively. In this way, we have the projection operator:
\begin{equation}
\Pi(M) =
\begin{cases*}
\frac{\lambda}{\twoinfnorm{M}} M, \ &\text{if}\ $\twoinfnorm{M} > \lambda$,\\
M,\ &\text{otherwise}.
\end{cases*}
\end{equation}
If we further pick the step size $\alpha_t$ via the Armijo rule~\cite{armijo1966minimization}, it can be shown that the sequence of $(U_t, V_t)$ will converge to a stationary point~\cite{bertsekas1999nonlinear}. The algorithm is summarized in Algorithm~\ref{alg:all}.

\begin{algorithm}[tb]
\caption{Max-norm Constrained 1-Bit Matrix Completion (MMC)}
\label{alg:all}
\begin{algorithmic}[1]
    \REQUIRE $Z \in \Rpn$ (observed samples), parameters $\lambda$, initial solution $(U_0, V_0)$, maximum iteration $\tau$.
    \ENSURE Optimal solution $(U^*, V^*)$.
    \FOR{$t=1$ to $\tau$}
      \STATE Compute the gradient by Eq.~\eqref{eq:gradient}:
       \begin{align*}
       U'_t &= \nabla_U^{} f(Z, U, V_{t-1}) \mid_{U=U_{t-1}},\\
       V'_t &= \nabla_V^{} f(Z, U_{t-1}, V) \mid_{V=V_{t-1}}.
       \end{align*}
      \STATE Compute the step size $\alpha_t$ according to Armijo rule.
      \STATE Compute the new iterate:
       \begin{align*}
       U_t &= \Pi(U_{t-1} - \alpha_t U'_t),\\
       V_t &= \Pi(V_{t-1} - \alpha_t V'_t).
       \end{align*}
    \ENDFOR
\end{algorithmic}
\end{algorithm}

The benefits of applying the factorization on the max-norm are two-folds: 1) the memory cost can be significantly reduced from $O(pn)$ of SDP to $O(d(p+n))$. 2) it facilitates the projected gradient algorithm, which is computationally efficient when working on large matrices (see Section~\ref{sec:exp}). However, note that Problem~\eqref{eq:uv prob} is non-convex. Fortunately,~\cite{burer2005local} proved that as long as we pick a sufficiently large value for $d$, then any local minimum of Eq.~\eqref{eq:uv prob} is a global optimum. In Section~\ref{sec:exp}, we will report the influence of $d$ on the performance. Actually, in Algorithm~\ref{alg:all}, the stopping criteria is set to be a maximum iteration. One may also check if it reaches a local minima as the stopping criteria, as discussed in~\cite{cai2013max}.

\subsection{Heuristic on $\lambda$}
The $\lambda$ is the only tunable parameter in our algorithm. For our problem, note that the data is of 1-bit measurements, {\em i.e.}, $ \lv Z_{ij} \rv = 1$ for $(i, j) \in \Omega$. Also note that $Z_{ij} = \bu(i) \bv(j)\trans$. Thus, $ \lv \bu(i) \bv(j)\trans \rv = 1$. So we have $\lambda \geq 1$. However, if we choose a large $\lambda$, the estimation $\lv X_{ij} \rv$ may deviate away from $1$. We find that $\lambda = 1.2$ lead to satisfactory improvement.


\section{Experiments}
\label{sec:exp}

In this section, we empirically evaluate our method for the matrix completion performance. We will first introduce the used datasets. In the experimental settings, we present the comparative methods and evaluation metrics. Then we report encouraging results on two benchmark datasets. We also examine the influence of matrix rank $d$.

\subsection{Datasets}  

We conduct the experiments on two benchmark datasets: Epinions and Slashdot. In these two datasets, the users are connected by explicit positive (trust) or negative (distrust) links ({\em i.e.,} the 1-bit measurements in $Z$). The first dataset contains 119,217 nodes (users) and 841,000 edges (links), 85.0\% of which are positive. The Slashdot dataset contains 82,144 users and 549,202 links, and 77.4\% of the edges are labeled as positive. Table \ref{dataSet} gives a summary description about the subset used in our experiment.

It is clear that the distribution of links are not uniform since each user has his/her individual preference and own friendship network. Following~\cite{huang2013social}, we select 2,000 users with the highest degrees from each dataset to form the observation matrix $Z$.

\begin{table}[h]
	\centering
	\caption{Description of 2 datasets}
	\label{dataSet}
	\centering
	\begin{tabular}{|l| r| r|}
		\hline
		Dataset & Epinions &Slashdot\\ \hline
		$\#$of Users  &2,000 &2,000 \\ \hline
		$\#$of Trust &171,731 &68,932  \\ \hline
		$\#$of Distrust &18,916 &20,032  \\ \hline
	\end{tabular}
\end{table}

\subsection{Experimental Settings}
\textbf{Baselines.} We choose four state-of-the-art methods as baselines, including SVP \cite{jain2010guaranteed}, SVT \cite{cai2010singular}, OPTSpace \cite{keshavan2010matrix} and RRMC \cite{huang2013social}. Since SVT and RRMC need a specified rank, we tune the rank for these methods and choose the best performance as the final result.

\textbf{Evaluation Metric.} Let $T$ be the index set of all observed entries. We use two evaluation metrics to measure the performance, mean average error (MAE) and root mean square error (RMSE), computed as follows:
\begin{equation*}
\begin{split}
MAE = & \sum _{(i,j)\in T \backslash \Omega} (X_{ij}-M_{ij})/(|T|-|\Omega|),\\
RMSE = & \sqrt{\sum _{(i,j)\in T \backslash \Omega} (X_{ij}-M_{ij})^2/(|T|-|\Omega|)},
\end{split}
\end{equation*}
where $|T|$ denotes the cardinality of $T$.

\begin{table*}[t]
	\centering
	\caption{MAE Results on Epinions Dataset }
	\label{maeEpin}
	\centering
	\begin{tabular}{|c| c| c| c| c| c|}
		\hline
		\multirow{2}*{Observed entries (\%)}  &\multicolumn{5}{c|} {Methods}\\
		\cline{2-6}&SVT &OPTSpace &SVP &RRMC &MMC\\ \hline
		10  &0.359$\pm$0.004		&0.289$\pm$0.019		&0.450$\pm$0.008		&0.576$\pm$0.001 & \textbf{0.254}$\pm$0.003 \\ \hline
		20 	&0.394$\pm$0.022	&0.236$\pm$0.005		&0.294$\pm$0.002		&0.518$\pm$0.002 & \textbf{0.212}$\pm$0.003 \\ \hline
		30  &0.360$\pm$0.057		&0.219$\pm$0.009		&0.248$\pm$0.001		&0.460$\pm$0.002 & \textbf{0.201}$\pm$0.002 \\ \hline
		40  &0.410$\pm$0.099		&0.205$\pm$0.008		&0.224$\pm$0.001		&0.418$\pm$0.002 & \textbf{0.193}$\pm$0.001\\ \hline
		50  &0.471$\pm$0.129		&0.197$\pm$0.007		&0.210$\pm$0.001		&0.386$\pm$0.001 & \textbf{0.190}$\pm$0.002 \\ \hline
		60  &0.476$\pm$0.146		&\textbf{0.197}$\pm$0.003		&0.199$\pm$0.001		&0.362$\pm$0.002& 0.206$\pm$0.003
		\\ \hline
	\end{tabular}
\end{table*}

\begin{table*}[t]
	\centering
	\caption{RMSE Results on Epinions Dataset }
	\label{rmseEpin}
	\centering
	\begin{tabular}{|c| c| c| c| c| c|}
		\hline
		\multirow{2}*{Observed entries (\%)}  &\multicolumn{5}{c|} {Methods}\\
		\cline{2-6}&SVT &OPTSpace &SVP &RRMC &MMC\\ \hline
		10  &0.513$\pm$0.010		&0.530$\pm$0.021		&0.610$\pm$0.010		&0.650$\pm$0.001 & \textbf{0.466}$\pm$0.004 \\ \hline
		20 	&0.559$\pm$0.031		&0.456$\pm$0.005		&0.459$\pm$0.002		&0.606$\pm$0.002 & \textbf{0.406}$\pm$0.004 \\ \hline
		30  &0.532$\pm$0.082		&0.422$\pm$0.011		&0.415$\pm$0.002		&0.563$\pm$0.002 & \textbf{0.383}$\pm$0.002 \\ \hline
		40  &0.620$\pm$0.171		&0.406$\pm$0.015		&0.394$\pm$0.002		&0.533$\pm$0.002 & \textbf{0.371}$\pm$0.002\\ \hline
		50  &0.719$\pm$0.225		&0.398$\pm$0.016		&0.381$\pm$0.001		&0.509$\pm$0.001 & \textbf{0.364}$\pm$0.001 \\ \hline
		60  &0.728$\pm$0.288		&0.403$\pm$0.009		&0.371$\pm$0.002	&	0.491$\pm$0.002 &\textbf{0.365}$\pm$0.002\\ \hline
	\end{tabular}
\end{table*}

\begin{table*}[t]
\centering
\caption{MAE Results on Slashdot Dataset }
\label{maeSlash}
\centering
\begin{tabular}{|c| c| c| c| c| c|}
\hline
\multirow{2}*{Observed entries (\%)}  &\multicolumn{5}{c|} {Methods}\\
\cline{2-6}&SVT &OPTSpace &SVP &RRMC &MMC\\ \hline
10  &0.679$\pm$0.008		&0.554$\pm$0.017		&0.755$\pm$0.005		&0.715$\pm$0.001 &\textbf{0.546}$\pm$0.008 \\ \hline
20 &0.562$\pm$0.004		&0.458$\pm$0.008		&0.582$\pm$0.007		&0.704$\pm$0.001 & \textbf{0.437}$\pm$0.008 \\ \hline
30  &0.513$\pm$0.030		&0.427$\pm$0.009		&0.501$\pm$0.003		&0.686$\pm$0.003 & \textbf{0.395}$\pm$0.006 \\ \hline
40  &0.506$\pm$0.041		&0.395$\pm$0.009		&0.460$\pm$0.002		&0.647$\pm$0.006 & \textbf{0.380}$\pm$0.009\\ \hline
50  &0.495$\pm$0.060		&0.376$\pm$0.004		&0.432$\pm$0.002		&0.609$\pm$0.002 & \textbf{0.366}$\pm$0.003 \\ \hline
60  &0.520$\pm$0.065		&0.362$\pm$0.011		&0.413$\pm$0.002		&0.585$\pm$0.002 & \textbf{0.350}$\pm$0.002\\ \hline
\end{tabular}
\end{table*}

\begin{table*}[t]
\centering
\caption{RMSE Results on Slashdot Dataset }
\label{rmseSlash}
\centering
\begin{tabular}{|c| c| c| c| c| c|}
\hline
\multirow{2}*{Observed entries (\%)}  &\multicolumn{5}{c|} {Methods}\\
\cline{2-6}&SVT &OPTSpace &SVP &RRMC &MMC\\ \hline
10  &0.788$\pm$0.008		&0.826$\pm$0.021		&0.873$\pm$0.004		&0.829$\pm$0.001 & \textbf{0.774}$\pm$0.008 \\ \hline
20 &0.718$\pm$0.020		&0.704$\pm$0.013		&0.746$\pm$0.007		&0.821$\pm$0.001 & \textbf{0.679}$\pm$0.009 \\ \hline
30  &0.680$\pm$0.043		&0.652$\pm$0.006		&0.680$\pm$0.003		&0.807$\pm$0.002 & \textbf{0.633}$\pm$0.008 \\ \hline
40  &0.670$\pm$0.056		&0.620$\pm$0.009		&0.647$\pm$0.003		&0.778$\pm$0.005 & \textbf{0.615}$\pm$0.011\\ \hline
50  &0.642$\pm$0.055		&0.596$\pm$0.009		&0.624$\pm$0.003		&0.749$\pm$0.002 & \textbf{0.581}$\pm$0.006 \\ \hline
60  &0.699$\pm$0.084		&0.577$\pm$0.011		&0.609$\pm$0.002		&0.730$\pm$0.002 & \textbf{0.566}$\pm$0.002\\ \hline
\end{tabular}
\end{table*}

\begin{table*}[!t]
	\centering
	\caption{The trade-off between accuracy and efficiency.}
	\label{time}
	\centering
	\begin{tabular}{|l| p{0.6cm}| p{0.8cm}| p{0.6cm}|p{0.6cm}| p{0.8cm}| p{0.6cm}| p{0.6cm}| p{0.8cm}| p{0.6cm}| p{0.6cm}| p{0.8cm}| p{0.6cm}| p{0.6cm}| p{0.8cm}| p{0.6cm}|}
		\hline
		\multirow{2}*{Dataset}  &\multicolumn{3}{c|} {SVT}  &\multicolumn{3}{c|} {OPTSpace}  &\multicolumn{3}{c|} {SVP} &\multicolumn{3}{c|} {RRMC}  &\multicolumn{3}{c|} {MMC}\\ 
		\cline{2-16}&MAE &RMSE &Time &MAE &RMSE &Time &MAE &RMSE &Time &MAE &RMSE &Time &MAE &RMSE &Time\\ \hline
		
		Epinions  &0.466 &0.618 &26.76  &0.288 &0.531 &15.26 &0.450 &0.610 &\textbf{0.84} &0.576 &0.651 &43.11 &\textbf{0.262} &\textbf{0.481} &1.92\\ \hline
		Slashdot &0.618 &0.728 &50.73 &0.458 &0.705 &9.23 &0.582 &0.746 &\textbf{0.83} &0.715 &0.830 &44.12&\textbf{0.437} &\textbf{0.679} &2.41 \\ \hline
	\end{tabular}
\end{table*}

\textbf{Training and Testing.} We randomly split the dataset for training and testing. In particular, the number of observation measurements $\Omega$ for training ranges from 10\% to 60\%, with step size 10\%. For each split, we run all the algorithms for 20 trials, with the training data in each trail being randomly sampled. Then, we report the mean and standard deviation of MAE and RMSE over all 20 trials.

\subsection{Experimental Results}
We report detailed results from Table~\ref{maeEpin} to Table~\ref{rmseSlash}. From the results in Tables \ref{maeEpin} and \ref{rmseEpin}, we observe that MMC outperforms the other methods in terms of both evaluation metrics on the Epinions dataset most of the time. In particular, when there are few observations 10 \% (which indicates a hard task), MMC obtains the RMSE of 0.466, much better than OPTSpace (0.530), SVP (0.610) and RRMC (0.650). Except on the case with 60\% observed entries, OPTSpace obtains the smallest MSE with 0.197, but our algorithm is comparative with 0.206. In a nut of shell, the gap between MMC and the baselines becomes larger as the fraction of observed entries decreases. 

Similarly, our method achieves the least MAE and RMSE on the Slashdot dataset (see Table \ref{maeSlash} and \ref{rmseSlash}). For instance with 30 \% observed entries, MMC obtains the MAE with less than 0.4, much better than the comparative methods, such as SVT (0.513), OPTSpace (0.427), SVP (0.501) and RRMC (0.686). In terms of RMSE, in the case of 20\% observed entries, the RMSE values of other methods are all above 0.7 while our method reaches 0.679. In sum, our method is superior than the comparative methods on two real-life datasets in terms of MSE and RMSE most of the time.

\begin{figure}[]
	\centering
	{\includegraphics[width=0.45\textwidth]{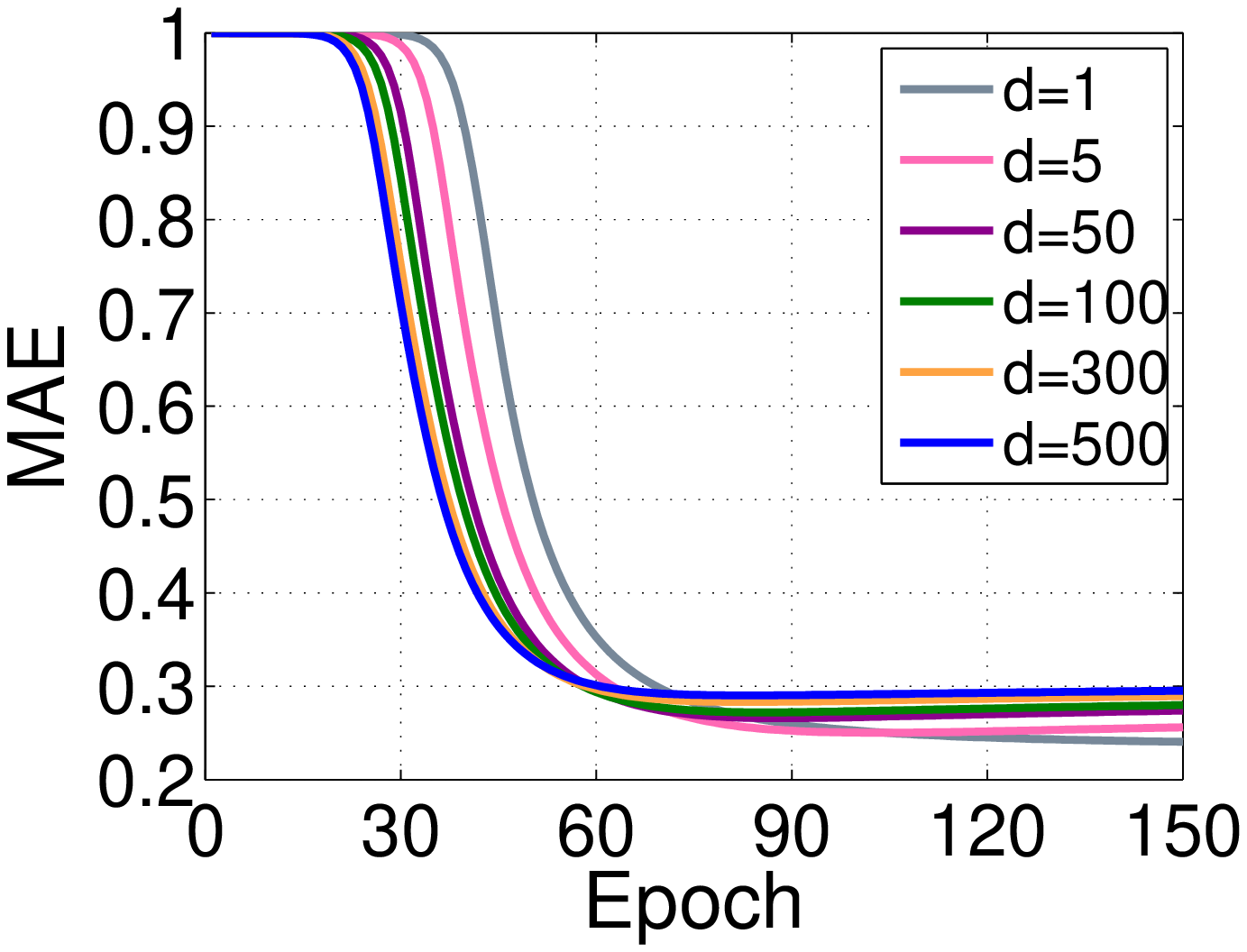}}
	 
	{\includegraphics[width=0.45\textwidth]{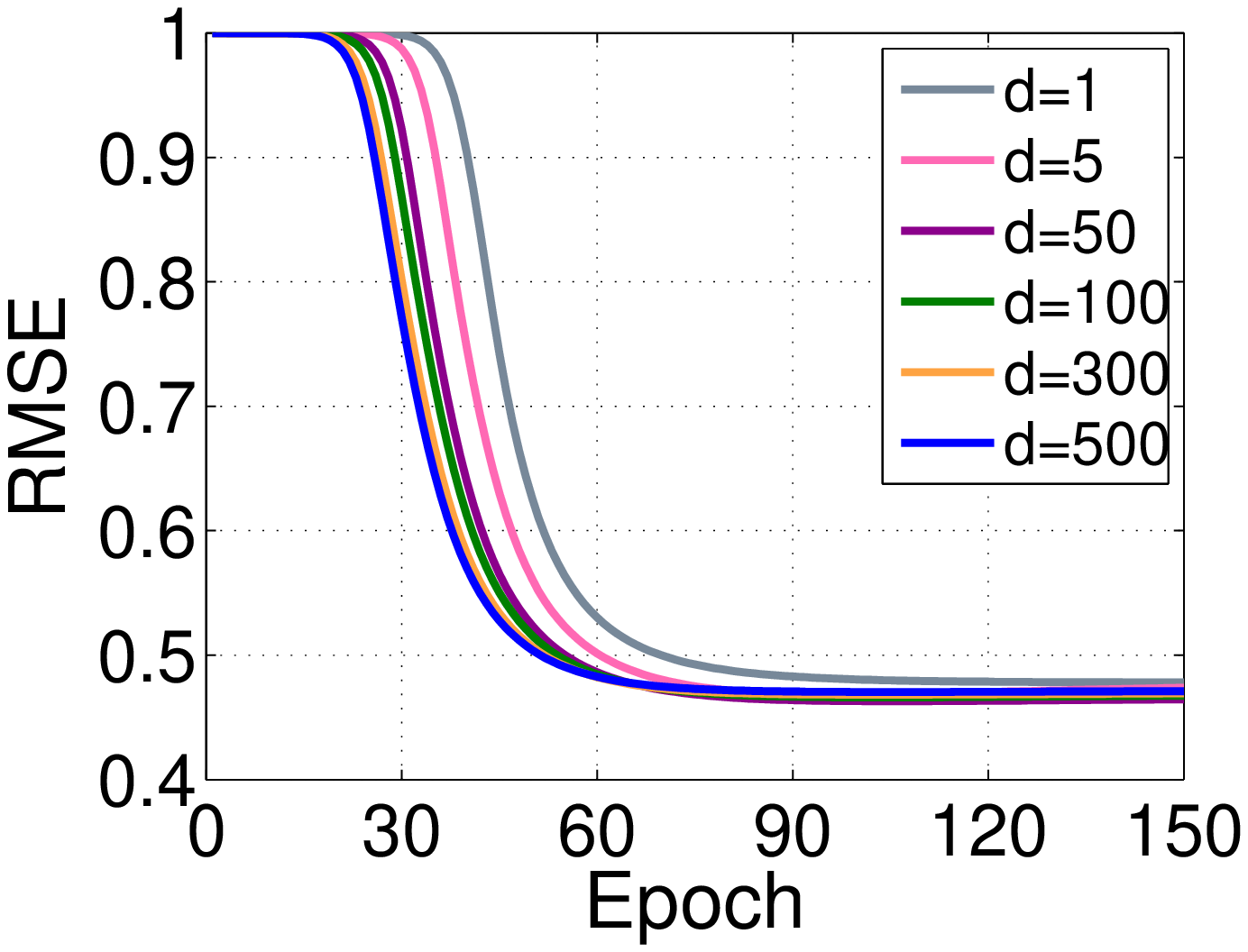}}
	
	\caption{MAE and RMSE in terms of different rank $d$ on the Epinions dataset.}
	\label{figrank}
	\vspace{-1em}
\end{figure}

Since we have studied the effectiveness of our method, here we examine the computational efficiency in Table~\ref{time}, which is important for practical applications. To test the time complexity of the methods, we report the averaged time cost on the Epinions dataset with 10\% observed entries and Slashdot with 20\% observed entries. To illustrate the trade-off between accuracy and efficiency, we also report the MAE and RMSE. As we see, SVP is the most efficient method, whose running time is 0.84 seconds on Epinions while ours is 1.92 seconds. On Slashdot, it also achieves the best performance in terms of efficiency. However, our method enjoys a significant improvement of MAE and RMSE compared to all baselines. Also, our algorithm is orders of magnitude faster than other three baselines. This implies that MMC favors a good trade-off between the accuracy and efficiency.

\subsection{Examine The Influence of $d$}  
The non-convex reformulation~\eqref{eq:uv prob} requires an explicit rank estimation $d$ on the true matrix. In this section, we investigate the influence of $d$ on the Epinions dataset as an example. The rank $d$ is chosen from [1, 5, 50 ,100, 300, 500] and the results are plot in Figure~\ref{figrank}. We observe that the rank has little influence on the performance. This is possibly because that the actual data has a low-rank structure (close to rank one). And from~\cite{burer2005local}, we know that if $d$ is larger than the actual rank, any local minimum of Eq.~\eqref{eq:uv prob} is also a global optima.

\section{Conclusion and Future Work}
\label{sec:conclusion}
In this paper, we formulated the social trust prediction in the matrix completion framework. In particular, due to the special structure of the social trust problem, {\em i.e.,} the measurements are 1-bit and the observed entries are non-uniformly sampled, we presented a max-norm constrained 1-bit matrix completion (MMC) algorithm. Since SDP solvers are not scalable to large scale matrices, we utilized a non-convex reformulation of the max-norm, which facilitates an efficient projected gradient decent algorithm. We empirically examined our algorithm on two benchmark datasets. Compared to other state-of-the-art matrix completion formulations, MMC consistently outperformed them, which meets with recently developed theories on max-norm. We also studied the trade-off between the accuracy and efficiency and observed that MMC achieved superior accuracy while keeping comparable computational efficiency.

The max-norm has been studied for several years and in many applications, such as collaborative filtering, clustering, subspace recovery. It is empirically and theoretically shown to be superior than the popular nuclear norm. This work investigates the power of max-norm for social trust problem and demonstrates encouraging results. It is interesting and promising to apply max-norm as a convex surrogate to other practical problems such as face recognition, subspace clustering etc.
\clearpage

\newpage
{\small
\bibliographystyle{named}
\bibliography{ijcai15}
}
\end{document}